\documentclass[preprint,12pt]{elsarticle}

\usepackage{graphicx}
\usepackage{amssymb}
\usepackage{amsmath}
\usepackage{algorithm}
\usepackage{algpseudocode}
\usepackage{enumerate}
\usepackage{color}

\newtheorem{theorem}{Theorem}
\newtheorem{lemma}[theorem]{Lemma}
\newtheorem{corollary}[theorem]{Corollary}

\begin{document}

\begin{frontmatter}



\title{Robust scheduling to minimize the weighted number of late jobs with interval due-date uncertainty}

\author{Maciej Drwal}

\address{Department of Computer Science\\
             Wroclaw University of Science and Technology,
              Wroclaw, Poland\\
              maciej.drwal@pwr.edu.pl
}

\begin{abstract}

We consider the class of single machine scheduling problems with the objective to minimize the weighted number of late jobs, under the assumption that completion due-dates are not known precisely at the time when decision-maker must provide a schedule. It is assumed that only the intervals to which the due-dates belong are known. The concept of maximum regret is used to define robust solution. A polynomial time algorithm is given for the case when weights of jobs are all equal. A mixed-integer linear programming formulation is provided for the general case, and computational experiments are reported.

\end{abstract}


\end{frontmatter}


\section{Introduction}

We consider the class of single machine scheduling problems for jobs with due-dates, with the objective to minimize the costs of missing deadlines. Problems of this type have been studied extensively in the literature \cite{brucker2007scheduling}, due to their broad practical applicability. The instances of these problems are defined by listing numerical parameters of each job: due-date, processing time, and, optionally, weight or cost. In practice, however, very often the exact values of these parameters are not known prior to the jobs' execution. At the time when decision-maker must provide a schedule they may be known only approximately. However, neglecting the inaccurate values of the parameters may lead to solutions that are unacceptably far away from the actual optimum \cite{ben2009robust}. This motivates the need for mathematical formulations that characterize optimal solutions which hedge against the uncertainty in the input data.

Two typical approaches used to handle uncertainty are stochastic optimization and robust optimization. In the former approach we need to assume or estimate a probability distribution of the input data. Then we usually require a solution to be feasible with a high probability, and that the value of the objective function, being a random variable, meets certain conditions \cite{heyman2003stochastic}. In the latter approach, which we consider in this paper, no probability distribution is used. Instead, we only define a set of possible parameter values of a problem instance. We require the solution to be unconditionally feasible in all scenarios -- realizations of the parameter values -- and that the solution is acceptable even in the worst-case realization.

An example application of the model studied in this paper is the task of preparing a schedule for a medical staff at a hospital. Suppose that a set of patients require urgent treatments, however, the specialized personnel and essential equipment can only be deployed to each case in a sequence. The success of treatment depends on whether it was applied to a patient on-time, but the exact due-date for a treatment to be effective is uncertain (usually only interval range can be designated). It is important to observe that in this type of critical applications it may not be advised to rely on stochastic modeling. Instead, a robust decision is needed, where the worst-case scenario is taken into consideration.

Similar situation is experienced whenever limited resources need to be deployed to tasks relying on the occurrence of specific events at uncertain time, be it in the supply chain management, delivery of perishable goods, responding to emergency calls, or hazardous environment exploration, to name a few.

A related criterion used in similar scheduling problems is the (weighted) tardiness \cite{brucker2007scheduling}, when the cost incurred by a late job is proportional to the time elapsed past the due-date. In this paper we assume instead that the costs of missing deadlines are constant. Our model applies to situations where occurrence of a specific event renders completing the job after its due-date irrelevant, i.e., the job is only useful if completed before deadline. Following the example of scheduling hospital staff, the weight of each job would indicate the risk level of not conducting the treatment on-time. The application of tardiness criterion would be adequate only for the type of medical conditions when risk levels increase continuously with time past some specific points.

In this paper we examine the setting when due-dates of jobs are given as intervals, with no additional probabilistic information available. 

In this approach the sense of an optimal solution is different from the one used in the traditional deterministic optimization. Although there are many kinds of robust formulations in optimization \cite{goerigk2016algorithm}, two typical robust objective functions used in discrete optimization are min-max and min-max regret \cite{kouvelis1997robust}. The {\it min-max} objective leads to solutions that have the smallest cost in the worst-case scenario (i.e., the scenario that maximizes the potential costs). However, in practice, the worst-case scenario may be unlikely to occur, thus this objective is sometimes too restrictive. The {\it min-max regret} objective relaxes the concept of min-max robustness in the following way: it measures how the value of a solution deviates from the optimal value computed for the worst-case scenario of the former solution. In other words, it is the maximum amount that is lost (or the additional cost that has to be paid) if the worst-case scenario had happened. 

Both the aforementioned robustness concepts can be viewed in the game theoretic terms: they represent a two-player game between an optimizing player and a malicious adversary, who controls the input data and decides upon the scenario after observing player's decisions.

In the presence of interval due-date uncertainty, the use of min-max criterion with the (weighted) number of late jobs leads to trivially considering only the scenario where all jobs assume their respective interval lower bounds. This may be argued to be overly pessimistic. But it is not the case when the min-max regret is utilized. The latter is also preferred as a robust criterion, as the solutions produced under this objective admit many important properties; the min-max regret satisfies a set of decision-theoretic axioms described in \cite{milnor1951games} (e.g., it is unique criterion to satisfy ordering, continuity, convexity, symmetry, column duplication, strong domination and strategic independence). Nevertheless, in some cases optimizing with min-max regret objective could be more difficult than with min-max. Experimental study indicates that this occurs in the general case of the considered problem. We show that if all jobs have the same weight then the complexity does not increase.

\subsection{Related Work}

Robust scheduling with interval parameter uncertainty has been considered in the works of Kouvelis and Yu \cite{kouvelis1997robust}. Since then, there were rather few results in this area.
A more recent and comprehensive review can be found in \cite{kasperskiminmax}, which discusses models for single machine with such criteria as: maximum weighted tardiness, weighted sum of completion times, as well as the problem of permutation flow shop and parallel machines scheduling problem with the maximum makespan criterion. Some results concerning the problem of min-max regret scheduling with total completion time criterion can be found in \cite{montemanni2007mixed}, \cite{lu2014robust}, \cite{lebedev2006complexity}, \cite{sotskov2009minimizing}, \cite{drwal2016complexity}. For the problem of min-max regret scheduling with maximum lateness criterion with interval-uncertain processing times a polynomial time algorithm is given in \cite{kasperski2005minimizing}. See also \cite{aissi2009min} for a survey on the subject of general robust combinatorial optimization.

For the scheduling problems with the criterion considered in this paper, most results were developed under a rather restrictive assumption that the uncertainty can be modeled by enumerating all possible scenarios (i.e., by discrete set of parameter values). In such discrete scenario case, the min-max scheduling to minimize the number of late jobs is known to be NP-hard even if job due-dates are deterministic and there are two processing time scenarios \cite{aloulou2008complexity}. This problem is strongly NP-hard and not approximable with a ratio less than 2 if all jobs have unit length processing times and the number of due-date scenarios is unbounded \cite{aissi2011minimizing}. The case of weighted number of late jobs of unit length is also NP-hard even for 2 weight scenarios and strongly NP-hard if the number of weight scenarios is unbounded. This problem (as well as its min-max regret variant) can be approximated within $K$ (the number of weight scenarios), and cannot be approximated within $O(\log K)$. If the number of due-date scenarios is constant, the unweighted min-max variant can be approximated with ratio 3 \cite{kasperskiminmax}.

In case of the interval representation of uncertainty, the variant with uncertain weights of jobs has been addressed in \cite{kasperski2008discrete}, and with uncertain processing times in \cite{drwal2017}. It is not known if there are polynomial time algorithms for such min-max regret scheduling problems, and no hardness results are known either. However, the case of interval-uncertain weights is a generalization of the \textsc{selecting items} problem, studied in \cite{averbakh2001complexity} and \cite{conde2004improved}. In the \textsc{selecting items} problem we want to pick $d$ items out of a set of $n$ items, so that the sum of their costs is minimized (the well-known \textsc{knapsack} problem is a generalization of the maximization version of \textsc{selecting items}, with arbitrary item sizes and designated capacity requirement, instead of a fixed number of $d$ items to pick). The \textsc{selecting items} problem is a special case of the considered scheduling problem, obtained when all due-dates are the same, and all jobs have unit processing time. Deterministic version of this problem can be simply solved by sorting, however this problem has interesting properties if the costs (weights) are uncertain. The \textsc{min-max regret selecting items} is one of very few known problems that are solvable in polynomial time when the uncertainty in the input data is represented by intervals. However, this problem is NP-hard in the discrete scenarios case, even for 2 scenarios. 

An interesting property of the problem variant considered in this paper is that the worst-case due-date scenarios are not necessarily the extreme scenarios (i.e., the ones consisting entirely of interval bounds). This, however, is the property of the interval-uncertain weights variant, and many other min-max (regret) discrete optimization problems.

\subsection{Contribution of the Paper}

Following the Graham classification scheme of scheduling problems \cite{brucker2007scheduling}, we denote the deterministic variant of the problem considered in this paper as $1|p_i=1|\sum w_iU_i$, whereas its uncertain counterpart will be prefixed by \textsc{interval-$d_i$ min-max regret} (abbreviated \textsc{i-$d_i$ mmr}). This notation indicates that the scheduling environment consists of a single machine, where duration of all jobs can be considered constant, and the quality assessment of a schedule involves the total weight of late jobs (implying a due-date $d_i$ attached to each job, see Section \ref{sec:form} for details).

The contributions of the paper are the following.
\begin{enumerate}[1)]
	\item It is shown how to compute the maximum regret for \textsc{i-$d_i$ mmr} $1|p_i=1|\sum w_iU_i$ for any given schedule (Section \ref{sec:max-regret}).
	\item A new mixed-integer linear program (MIP) is developed for the general case of the problem (Section \ref{sec:mip}).
	\item A polynomial time algorithm is given for the special case with unit weights, i.e., problem \textsc{i-$d_i$ mmr} $1|p_i=1|\sum U_i$, implying that in this case the uncertain counterpart is not more difficult to solve that the deterministic problem (Section \ref{sec:unit}).
	\item Computational results are reported from the application of MIP using state-of-the-art software solver. Obtained results are compared against three heuristic solution methods: decomposition algorithm, lower bound heuristic and mid-point scenario heuristic (Section \ref{sec:experiment}).
\end{enumerate}

The remaining part of the paper is organized as follows. A formulation of the considered problem is stated in Section \ref{sec:form}. The main results of the paper are given in Sections \ref{sec:mip} (MIP for the general case) and Section \ref{sec:unit} (a simple algorithm for special case). Subsequently, practical applicability of the MIP is examined in Section \ref{sec:experiment}, where numerical study is carried out. Finally, Section \ref{sec:conclusions} concludes the paper.

\section{Problem Formulation}\label{sec:form}

\subsection{Definition of Nominal Problem}

We start by defining a general scheduling problem with weighted number of late jobs criterion. The full list of symbols used in definitions can be found in Table \ref{tab:0}. 

Consider the scheduling problem $1|p_i=1|\sum w_i U_i$, which we will henceforth refer to as the {\em nominal problem}. Given is the set $J = \{ 1, 2, \ldots, n \}$ of unit-length jobs, each described by two parameters: $d_i$, a due-date (deadline), that is, the time by which $i$th job must be completed; and $w_i$, the weight (or cost) of $i$th job. A solution is any permutation of job indices $\pi = (\pi(0), \pi(1), \ldots, \pi({n-1}))$, where $\pi(j)$ is the index of job to be executed at time $j$. We define by $\pi^{-1}(i)$ the time when job $i$ starts in the schedule $\pi$ (that is, $\pi^{-1}(i) = j$ if and only if $\pi(j) = i$). By $\mathcal{P}$ we denote the set of all permutations of $\{ 1,\ldots, n \}$. 

Let $C(\pi, i)$ denote the completion time of job $i \in J$ in the schedule $\pi$, that is:
$$
C(\pi, i) = \sum_{j: \; 0 \leq j \leq \pi^{-1}(i)} p_{\pi(j)}.
$$
If a job is completed after its due-date, a penalty equal to the weight $w_i$ is incurred. The objective is to find a schedule that minimizes the sum of weights of the jobs that complete after their deadlines, i.e., 
\begin{equation}\label{obj-fun}
	F(\pi) = \sum_{i \in J} w_i U_i(\pi),
\end{equation}
where:
\begin{equation}\label{Ui}
	U_i(\pi) = \left\{ \begin{array}{ll}
			0, & C(\pi, i) \leq d_i, \\
			1, & \textrm{otherwise}.
		  \end{array} \right.
\end{equation}
If a job is scheduled before its due-date, we say that the job is {\it on-time}. Otherwise, we say that the job is {\it late}.

Note that the nominal problem has the structure of a matroid, and can be solved in polynomial time using a greedy algorithm \cite{cormen2001}. The subsets of jobs that can be on-time in some schedule form independent sets of the matroid. Optimal solutions correspond to maximal independent sets with the largest total weight. Any feasible solution can be transformed into a cost-equivalent solution in the canonical form, where all on-time jobs are scheduled before all late jobs, and the on-time jobs appear in the schedule in the order of nondecreasing due-dates.

\subsection{Definition of Uncertain Problem}

In this variant of $1|p_i=1|\sum w_iU_i$ problem, the due-dates are uncertain, and given are the interval due-dates $[d_i^-, d_i^+]$ for each job $i \in J$, where $J = \{ 1, 2, \ldots, n \}$ is the set of given jobs. We assume that all weights $w_i$ are certain. We denote $W = \sum_{i \in J} w_i$. Moreover, we assume that all the numeric data describing a problem instance are positive integers (the realizations of the due-dates may be arbitrary real numbers within the considered intervals, although bounding values $d_i^-$ and $d_i^+$ are always integers).

We consider the \textsc{interval-$d_i$ min-max regret} $1|p_i=1|\sum w_i U_i$ variant of the problem. Let $\mathcal{U} = \{ (d_1, \ldots, d_n) : d_i^- \leq d_i \leq d_i^+ \}$ be the set of all possible realizations of due-dates. We call $S \in \mathcal{U}$ a {\it scenario}. We will write $S = \{ d_i^S \}_{i=1}^n$ to denote all jobs' due-dates in a scenario $S$. Let $F(\pi, S)$ be the value of the nominal objective function \eqref{obj-fun} computed for a scenario $S \in \mathcal{U}$, and $F^*(S)$ be the value of an optimal solution for a scenario $S \in \mathcal{U}$. Note that this time the terms \eqref{Ui} in the objective function depend on the scenario.

We define the {\it regret} $R(\pi, S)$ of a solution $\pi$ in a scenario $S$:
\begin{equation}\label{def:regret}
	R(\pi, S) = F(\pi, S) - F^*(S).
\end{equation}
A scenario $S$ that maximizes $R(\pi, S)$ is called a {\it worst-case scenario}. A schedule $\sigma(S)$ that is optimal for this scenario is called a {\it worst-case alternative} for solution $\pi$.

In the interval min-max regret version of the problem the goal is to schedule the jobs in such an order that minimizes the maximum regret of the nominal problem's objective; that is:
\begin{equation}\label{def:min-max-regret}
    \min_{\pi \in \mathcal{P}} Z(\pi) = \min_{\pi \in \mathcal{P}} \max_{S \in \mathcal{U}} R(\pi, S) = \min_{\pi \in \mathcal{P}} \max_{S \in \mathcal{U}} \left( F(\pi, S) - F^*(S) \right).
\end{equation}
A solution that minimizes the regret is called {\em optimal robust} solution.

It is convenient to use terminology from game theory in order to describe this problem. Any min-max regret problem can be seen as a two-player game in which the minimizing player controls solution $\pi$, and the maximizing player controls the scenario $S$. The minimizing player (further called simply {\em player}) makes the decision first. The maximizing player (further called {\em adversary}) observes this decision, and then selects a scenario $S$. The computation of game's outcome involves solving the nominal (deterministic) problem for the worst-case scenario. The problem of computing the pair $(S, \sigma)$ (i.e., the worst-case scenario and the worst-case alternative solution) for a given solution $\pi$ is called the {\em adversarial problem}.

\subsection{Computation of Maximum Regret}\label{sec:max-regret}

Given a schedule $\pi \in \mathcal{P}$ for the problem $1|p_i=1|\sum w_i U_i$, the maximum regret can be expressed as:
\begin{equation}\label{immr-1pwUd-obj}
	Z(\pi) = \max_{S \in \mathcal{U}, \, \sigma \in \mathcal{P}} \left( 
	                                      \sum_{\substack{k: \, \pi^{-1}(k) \geq d_k, \\
										                        \sigma^{-1}(k) < d_k}} w_k - 
	                                      \sum_{\substack{k: \, \pi^{-1}(k) < d_k, \\
										                        \sigma^{-1}(k) \geq d_k }} w_k
	                                \right),
\end{equation}
where $S = \{ d_i \}_{i=1}^n$. To see this, observe that if a job is on-time in both $\pi$ and $\sigma$, the corresponding weight does not appear in the objective function of the nominal problem. If a job is late in $\pi$, then its cost is included in the first sum, provided that the job is on-time in $\sigma$. Similarly, if a job is late in $\sigma$, then its cost is included in the second sum, provided that the job is on-time in $\pi$.

For a given schedule $\pi$, the worst-case scenario can be computed as follows. 
The worst-case due-date $d_i^S$ of job $i \in J$ is:
\begin{enumerate}[(i)]
	\item if $\pi^{-1}(i) < d_i^-$ then $d_i^S = d_i^+$ (job $i$ is always on-time, regardless of the scenario),\label{wc-1}

	\item if $\pi^{-1}(i) \geq d_i^+$ then $d_i^S = d_i^+$ (job $i$ is always late, regardless of the scenario),\label{wc-2}

	\item if $d_i^- \leq \pi^{-1}(i) < d_i^+$ then $d_i^S = \pi^{-1}(i)$ (job $i$ may be on-time or late, depending on the scenario).\label{wc-3}
\end{enumerate}

To obtain the worst-case alternative for $\pi$ it is enough to run the greedy algorithm and solve the nominal problem with the obtained weights $d_i^S$, $i \in J$.

In cases \eqref{wc-1} and \eqref{wc-2} the choice of the scenario does not have any effect from the perspective of the player, and the adversary may choose the most favorable values $d_i^+$. Such choice cannot decrease the regret as compared to any other choice $d_i < d_i^+$, since the adversary would be able to insert on-time the largest number of jobs.

In case \eqref{wc-3} the choice of $d_i^S=\pi^{-1}(i)$ causes the job to be late in the player's schedule $\pi$. This adds $w_i$ to the value of the regret if the job $i$ is on-time in the adversary's schedule $\sigma$, or leaves the value of the regret unchanged otherwise. When $d_i > \pi^{-1}(i)$ then job $i$ becomes on-time in $\pi$, thus the adversary has nothing to gain by increasing $d_i^S$ above $\pi^{-1}(i)$. When $d_i^S < \pi^{-1}(i)$ the player still misses the due-date with job $i$, but the number of jobs that can be inserted on-time by the adversary in $\sigma$ is not necessarily maximal. Consequently, the choice $d_i^S = \pi^{-1}(i)$ is the most advantageous for the adversary.

In the worst-case scenario, all the jobs scheduled after their due-dates' lower bounds would be late in the player's schedule. Observe that the worst-case scenario may consist of arbitrary integer values within uncertainty intervals $[d_i^-, d_i^+]$.

\section{Mixed-integer Linear Programming Formulation}\label{sec:mip}

In this section we present a mixed-integer linear program (MIP) for computing min-max regret schedule of the nominal problem with weighted number of late jobs criterion and uncertain due-dates. This program allows for efficiently solving the problem instances of moderate size, and obtaining good approximate solutions of larger instances, using standard branch-and-cut type of algorithms. 

Although the problem of minimizing the maximum regret \eqref{def:min-max-regret} involves a nested maximization subproblem, an efficient reformulation is possible due to the use of the dual of nominal problem.

Considering known due-dates, indexed so that $d_1 \leq d_2 \leq \ldots \leq d_n$, the nominal problem can be formulated as the following linear program (LP):

\begin{equation}
	\textrm{maximize } \sum_{k \in J} \hat{x}_k w_k, \label{lp-1:obj}
\end{equation}
subject to:
\begin{align}
	\forall_{k \in J} \;\; \sum_{i=1}^k \hat{x}_i \leq d_k, \label{lp-1:1}
\\
	\forall_{k \in J} \;\; 0 \leq \hat{x}_k \leq 1. \label{lp-1:2}
\end{align}
Note that the value of the objective function \eqref{obj-fun} can be computed as $W - \sum_{k \in J} \hat{x}^*_k w_k$, where $\hat{\bf x}^*$ is an optimal solution of \eqref{lp-1:obj}--\eqref{lp-1:2}.
It can be seen that the constraint matrix of \eqref{lp-1:1}--\eqref{lp-1:2} is unimodular, thus optimal solutions must consist of all integer values. Decision variable $\hat{x}_i$ assumes value $1$ if and only if $i$th job is scheduled on-time. The dual of this LP is:
\begin{equation}
	\textrm{minimize } \sum_{k \in J} \nu_k + \sum_{j \in J} d_j \lambda_j, \label{lp-2:obj}
\end{equation}
subject to:
\begin{align}
	\forall_{k \in J} \;\; w_k \leq \nu_k + \sum_{i=k}^n \lambda_i, \label{lp-2:1}
\\
	\forall_{k \in J} \;\; \nu_k, \lambda_k \geq 0. \label{lp-2:2}
\end{align}

Let us rewrite the problem of minimizing \eqref{immr-1pwUd-obj} as:
$$
	Z^* = \min_{\pi \in \mathcal{P}} Z(\pi) = \min_{\pi \in \mathcal{P}} \max_{S \in \mathcal{U}} \left( F(\pi, S) - \max_{\sigma  \in \mathcal{P}} F(\sigma, S) \right).
$$
As it has been shown in Section \ref{sec:max-regret}, the maximization over scenarios can be computed directly given the schedule $\pi$. Moreover, the inner maximization problem (i.e., maximization over schedules $\sigma$) can be substituted by the dual problem \eqref{lp-2:obj}--\eqref{lp-2:2}, that is, minimization over $\nu_k, \lambda_k$, for $k \in J$. We obtain the following program:

\begin{equation}
	\textrm{minimize } \sum_{k \in J} y_k w_k + \sum_{k \in J} \nu_k + \sum_{k \in J} k \lambda_k - W, \label{mip-1:obj}
\end{equation}
subject to:
\begin{align}
	\forall_{k \in J} \forall_{j < d_k^- \; \vee \; j \geq d_k^+} & \;\;\; x_{kj} \leq \delta_{k{d_k^+}}, \label{mip-1:1}
\\
	\forall_{k \in J} \forall_{d_k^- \leq j < d_k^+} & \;\;\; x_{kj} \leq \delta_{kj}, \label{mip-1:2}
\\
    \forall_{k \in J} & \;\;\; \sum_{j \in J} j x_{kj} - \sum_{j \in J} j \delta_{kj} +1 \leq n y_k, \label{mip-1:3}
\\
	\forall_{k \in J} \forall_{j \in J} & \;\;\; D_{kj} = \sum_{i=1}^j \delta_{ki}, \label{mip-1:4}
\\
	\forall_{k \in J} & \;\;\; w_k \leq \nu_k + \sum_{j \in J} z_{kj}, \label{mip-1:5}
\\
	\forall_{k \in J} & \;\;\; \sum_{j \in J} x_{kj} = 1, \label{mip-1:6}
\\
	\forall_{j \in J} & \;\;\; \sum_{k \in J} x_{kj} = 1, \label{mip-1:7}
\\
	\forall_{k \in J} & \;\;\; \sum_{j \in J} \delta_{kj} = 1, \label{mip-1:8}
\\
	\forall_{k \in J} \forall_{j \in J} & \;\;\; z_{kj} - W D_{kj} \leq 0, \label{mip-1:9}
\\
	\forall_{k \in J} \forall_{j \in J} & \;\;\; \lambda_j + W D_{kj} - z_{kj} \leq W,
\\
	\forall_{k \in J} \forall_{j \in J} & \;\;\; z_{kj} - \lambda_j \leq 0, \label{mip-1:11}
\\
	\forall_{k \in J} \forall_{j \in J} & \;\;\; D_{kj} \geq 0, \lambda_j \geq 0, \nu_j \geq 0, z_{kj} \geq 0,
\\
	\forall_{k \in J} \forall_{j \in J} & \;\;\; x_{kj} \in \{ 0, 1 \}, \delta_{kj} \in \{ 0, 1 \}, y_k \in \{ 0, 1 \}. \label{mip-1:13}
\end{align}

Binary variables $x_{kj}$, $k,j \in J$, encode permutation $\pi \in \mathcal{P}$, which is guaranteed by the set of constraints \eqref{mip-1:6}--\eqref{mip-1:7}. Binary variables $\delta_{kj}$ encode the worst-case due-dates, that is, $\delta_{kj}=1$ if and only if $k$th job has its worst-case due-date equal to $j$. The semi-assignment constraints \eqref{mip-1:8} guarantee that each job has a well-defined due-date. The constraints \eqref{mip-1:1}--\eqref{mip-1:2} define the values of worst-case due-dates, based on permutation encoded by $x_{kj}$, as shown in Section \ref{sec:max-regret}; that is, constraints \eqref{mip-1:1} correspond to the cases (i) and (ii), when job is scheduled either before its earliest possible due-date, or after its latest possible due-date, while constraints \eqref{mip-1:2} correspond to the case (iii), when job is scheduled at time within its uncertainty interval. Since for each $k \in J$ there is exactly one $j$, such that $x_{kj} = 1$, thus exactly one $\delta_{kj}$ must be set to $1$ accordingly. 

Binary variables $y_k$, $k \in J$, indicate that $k$th job is late in the schedule (when $y_k=1$), thus their cost is counted in the objective function. The constraints \eqref{mip-1:3} are used to determine which jobs are late in the solution, by subtracting the position index of $k$th job from its due-date value, indicated by $j \delta_{kj}$. If the resulting value is $-1$ or less, then $k$th job completes on-time in the schedule. Since in this case the lefthand side of the constraint is nonpositive, the corresponding variable $y_k$ would assume the value zero. Otherwise, when $k$th job is late in the schedule, the lefthand side of the constraint must be positive, thus $y_k$ must be positive.

Constraints \eqref{mip-1:4}--\eqref{mip-1:5} correspond to the dual constraints \eqref{lp-2:1} for the adversarial subproblem. Constraints \eqref{mip-1:5} could be rewritten using mixed-quadratic terms $D_{kj} \lambda_j = z_{kj}$, for all $k,j \in J$, as:
\begin{equation}\label{mip-1:aux1}
w_k \leq \nu_k + \sum_{j \in J} D_{kj} \lambda_j.
\end{equation}
These terms are linearized in a standard way using variables $z_{kj}$ and the set of constraints \eqref{mip-1:9}--\eqref{mip-1:11}. The inequality \eqref{mip-1:aux1} is obtained as follows. Since the constraint \eqref{lp-1:1} of the nominal problem LP requires due-dates to be sorted in nondecreasing order, we can write it as:
\begin{equation}\label{mip-1:aux2}
	\forall_{j = 1,2,\ldots,n} \;\;\; \sum_{i=1}^j \sum_{k=1}^n \delta_{ki} \hat{x}_k \leq j,
\end{equation}
where $\hat{x}_k$ is a primal variable of the nominal problem, denoting whether $k$th job is on-time. On the lefthand side of the above inequality, included are only these variables $\hat{x}_k$, which correspond to the jobs $k$ with due-dates less than or equal to $j$, for each $j=1,2,\ldots,n$. Consequently, the variables $D_{kj} = \sum_{i=1}^j \delta_{ki}$ are introduced (constraints \eqref{mip-1:4}) in order to control the placement of variables $\hat{x}_k$ in appropriate constraints of this type. This is necessary, since the order of worst-case due-dates, indicated by the values of $\delta_{ki}$, depends on the solution $\pi$. Adversarial subproblem's primal constraints \eqref{mip-1:aux2} correspond to the dual variables $\lambda_j$, while the constraints $0 \leq \hat{x}_k \leq 1$ correspond to the dual variables $\nu_k$.

\section{Solution Algorithm for Unit-Weight Case}\label{sec:unit}

The mixed-integer formulation of the problem with arbitrary weights leads to the solution algorithm with exponential time complexity in the worst case. It is unknown whether there exists a polynomial time algorithm for that problem. It can be shown, however, that restricted variants of this problem admit efficient solution algorithms.

We now consider a special case of the considered scheduling problem, where all weights are equal to 1. In this variant, we are interested only in minimizing the {\it number} of late jobs (equally penalizing each missed due-date). We will show that this problem can be solved in polynomial time.

We start from an observation that in a robust solution, the jobs that are placed on positions within their uncertainty intervals must be scheduled in an order of non-increasing upper bounds of their due-date intervals $d_i^+$.

For a given schedule $\pi$ denote the set $L(\pi) = \{ j \in J : \; \pi^{-1}(j) \geq d_j^- \}$ of jobs scheduled no earlier than their earliest possible due-date.


\begin{lemma}\label{lem:1}	
	Let $\pi'$ be a schedule, and let $\{ d_j^{S'} \}_{j \in J}$ be the set of worst-case due-dates for the schedule $\pi'$. There exists a schedule $\pi$ satisfying:
	\begin{equation}\label{lem:1:eq1}
	\forall{i,j \in L(\pi)} \;\;\; \pi^{-1}(i) < \pi^{-1}(j) \Rightarrow d_i^+ \geq d_j^+,
	\end{equation}
	and a bijection $b : J \rightarrow J$, such that:
	\begin{equation}\label{lem:1:eq2}
	\forall{j \in L(\pi)} \;\;\; d_j^S \leq d_{b(j)}^{S'},
	\end{equation}
	where $\{ d_j^S \}_{j \in J}$ is the set of worst-case due-dates for the schedule $\pi$.
	
\end{lemma}

{\em Proof.}
We show that given arbitrary schedule $\pi'$, it is always possible to construct a schedule $\pi$ for which \eqref{lem:1:eq1} is satisfied by performing a finite number of exchanges of pairs of jobs. Let us assume that $\pi'$ does not satisfy the corresponding property \eqref{lem:1:eq1}, as otherwise the claim is trivially true.

Starting with the schedule $\pi_1 = \pi'$, we take any pair of jobs $i,j \in L(\pi_1)$, such that $\pi_1^{-1}(i) < \pi_1^{-1}(j)$ and $d_i^+ < d_j^+$ and swap their positions. This operation is repeated until \eqref{lem:1:eq1} is satisfied for the modified schedule. We will show how to define a bijection $b$ for which the statement \eqref{lem:1:eq2} is satisfied after every such exchange operation. This implies that the statement is satisfied by $\pi$ and for any given $\pi'$.

Now we describe a single exchange operation. This operation transforms a schedule $\pi_1$ into a schedule $\pi_2$. We denote the worst-case scenario $S_1$ of the schedule $\pi_1$ before the exchange operation, and the worst-case scenario $S_2$ of the schedule $\pi_2$ after the exchange. Let $i,j \in L(\pi_1)$ be the indices selected for swapping, i.e., $k = \pi_1^{-1}(i) < \pi_1^{-1}(j) = l$ and $d_i^+ < d_j^+$. In the obtained schedule $\pi_2$ we have $\pi_2(k) = j$ and $\pi_2(l) = i$. We maintain a bijection $b$, initialized to identity (i.e., $b(i)=i$ for all $i \in J$), which will be modified appropriately after each exchange operation.

We can also assume that $d_j^- \leq k$. Otherwise we have $d_j^{S_2} = d_j^+$ (job $j$ is on-time regardless of the scenario, see case (i) in Section \ref{sec:max-regret}), and since job $j \notin L(\pi_2)$, we only need to verify whether $d_i^{S_2} \leq d_{b(i)}^{S_1}$. Observe that since $i \in L(\pi_1)$, it follows that $i \in L(\pi_2)$.

We compute the worst-case due-dates $d_{i}^{S_1} = \min \{ k, d_{i}^+ \}$, $d_{j}^{S_1} = \min \{ l, d_{j}^+ \}$, $d_{i}^{S_2} = \min \{ l, d_{i}^+ \}$, $d_{j}^{S_2} = \min \{ k, d_{j}^+ \}$ in all possible ordering sequences satisfying $k < l$ and $d_{i}^+ \leq d_{j}^+$. There are six cases:

\begin{enumerate}[(i)]
	\item $k < l \leq d_i^+ \leq d_j+$, where $l = d_i^{S_2} = d_j^{S_1} = l$, and $k = d_j^{S_2} = d_i^{S_1} = k$, thus we put $b(i) = j$, and $b(j) = i$,
	\item $k \leq d_i^+ \leq l \leq d_j^+$, where $d_i^+ = d_i^{S_2} \leq d_j^{S_1} = l$, and $k = d_j^{S_2} = d_i^{S_1} = k$, thus we put $b(i) = j$, and $b(j) = i$,
	\item $k \leq d_i^+ \leq d_j^+ \leq l$, where $d_i^+ = d_i^{S_2} \leq d_j^{S_1} = d_j^+$, and $k = d_j^{S_2} = d_i^{S_1} = k$, thus we put $b(i) = j$, and $b(j) = i$,
	\item $d_i^+ \leq k < l \leq d_j^+$, where $d_i^+ = d_i^{S_2} = d_i^{S_1} = d_i^+$, and $k = d_j^{S_2} \leq d_j^{S_1} = d_j^+$, thus we put $b(i) = i$, and $b(j) = j$,
	\item $d_i^+ \leq k \leq d_j^+ \leq l$, where $d_i^+ = d_i^{S_2} = d_i^{S_1} = d_i^+$, and $k = d_j^{S_2} \leq d_j^{S_1} = d_j^+$, thus we put $b(i) = i$, and $b(j) = j$,
	\item $d_i^+ \leq d_j^+ \leq k < l$, where $d_i^+ = d_i^{S_2} = d_i^{S_1} = d_i^+$, and $d_j^+ = d_j^{S_2} = d_j^{S_1} = d_j^+$, thus we put $b(i) = i$, and $b(j) = j$.
\end{enumerate}

It follows that after each exchange operation we have a schedule $\pi_2$ and a bijective function $b$, such that for all $i,j \in L(\pi_2)$ we have $d_i^{S_2} \leq d_{b(i)}^{S_1}$ and $d_j^{S_2} \leq d_{b(j)}^{S_2}$. This completes the proof.
\qed

\bigskip

Denote by $I(\pi)$ the maximal independent set of jobs in the worst-case scenario for the schedule $\pi$. Recall that an independent set of a scheduling matroid corresponds to the set of jobs that can be scheduled on-time simultaneously. A maximal independent set has the property that no superset of this set is independent. Since the objective of the considered problem is the number of late jobs, optimal solutions correspond to the maximal independent sets.

Lemma \ref{lem:1} implies the following fact:

\begin{corollary}\label{cor:1}
	
	Let $\pi$ be a schedule satisfying \eqref{lem:1:eq1}. Consider any other schedule $\pi'$, such that $L(\pi') = L(\pi) = L$. Then $|I(\pi)| \leq |I(\pi')|$.

\end{corollary}

{\em Proof.}
Denote by $\{ d_j^{S} \}_{j \in J}$ the set of worst-case due-dates for schedule $\pi$, and by $\{ d_j^{S'} \}_{j \in J}$ the set of worst-case due-dates for schedule $\pi'$. Since $I(\pi)$ is an independent set of matroid, we have for $t = 1, 2, \ldots, |L|$:
$$
| \{ j \in I(\pi) : \; d_j^S \leq t \} | \leq t.
$$

From Lemma \ref{lem:1} we have $d_j^{S} \leq d_{b(j)}^{S'}$ for all $j \in L$. Thus:
$$
t \geq | \{ j \in I(\pi) : \; d_j^S \leq t \} | \geq | \{ j \in I(\pi) : \; d_{b(j)}^{S'} \leq t \} |.
$$
Thus $I(\pi)$ is also an independent set for the set of jobs with due-dates $\{ d_j^{S'} \}_{j \in J}$. Consequently, $I(\pi) \subseteq I({\pi'})$, where $I(\pi')$ is the maximal independent set for these due-dates. The claim follows.
\qed

\bigskip

Since the jobs executed at or after their lower bounds of due-date intervals are late in the player's schedule, it is beneficial for the player to schedule them so that their worst-case due-dates assume small values. This would force the adversary to schedule the least number of these jobs on-time. The adversary can schedule on-time all the jobs from the maximal independent set $I(\pi)$. The Corollary \ref{cor:1} states that when the player schedules the late jobs in the order of non-increasing due-dates $d_j^+$, the size of a maximal independent set of these jobs is the least possible.

Based on this observation, a simple algorithm for computing an optimal robust solution can be given.

\medskip

{\bf Algorithm A.}
\begin{enumerate}
	\item Solve an instance of an auxiliary nominal problem $1|p_i=1|\sum w_iU_i$ with the set of weights $\tilde{w}_j = -d_j^+$, $j \in J$, and the set of due-dates $\tilde{d}_j = d_j^-$, $j \in J$. Denote by $I$ the set of on-time jobs.
	\item Schedule each job $j \in I$ , so that $\pi^{-1}(j) < d_j^-$. 
	\item Schedule the set of jobs $I' = \{ j \in J: \; j \notin I \}$ in the order of non-increasing values $d_j^+$ on positions $|I|, |I|+1, \ldots, n$.
\end{enumerate}

Step 1 of this algorithm can be completed in time $O(n^2)$ with the use of greedy algorithm: start with an empty set $I$, sort the jobs in the order of nondecreasing upper bounds of due-date intervals: $d_1^+ \leq d_2^+ \leq \ldots \leq d_{|I|}^+$, then, proceeding in that order, add job $j$ to the set $I$ if:
$$\forall_{t = 1, \ldots, j} \;\; | \{ i \in I \cup \{ j \} : \; d_i^- \leq t \}| \leq t.$$
It is always possible to schedule jobs $j \in I$ before their lower bounds of due-date intervals (step 2), since $I$ is an independent set for these due-dates. Step 3 of the algorithm can be completed in time $O(n \log{n})$.

\begin{theorem}
	Algorithm A outputs a robust optimal solution $\pi$ for \textsc{interval-$d_i$ min-max regret} $1|p_i=1|\sum U_i$ problem with uncertain due-dates.
\end{theorem}

{\em Proof.}
We show that Algorithm A computes an optimal robust schedule. The maximum regret of a solution $\pi$ can be expressed as
$$Z(\pi) = n_a(\pi) - n_p(\pi),$$
where $n_p(\pi)$ is the number of on-time jobs in $\pi$ in the worst-case scenario, and $n_a(\pi)$ is the number of on-time jobs in the worst-case alternative schedule (the adversary's schedule). The correctness of the algorithm follows from an observation that the player should always schedule on-time the jobs from maximal independent set $I$ of jobs with lower bounds of due-date intervals $d_j^-$ (i.e., as many jobs as possible that would be on-time regardless of a scenario). 

Suppose the contrary, that a job $j$ is late in the player's schedule, but could have been on-time in the worst-case scenario, if it were placed at position $k < d_j^-$. Let $l$ be the position of job $j$ in the schedule. The worst-case due-date of $j$ is equal to $l$, if $l > d_j^+$, or is equal to $d_j^+$, otherwise. From Corollary \ref{cor:1}, all the jobs scheduled not earlier than their lower bounds of due-date interval must be placed in the order of non-increasing upper bounds of due-date interval, in order to keep the number $n_a(\pi)$ the smallest possible (for a particular set of late jobs in $\pi$). Consequently, if job $j$ were moved from position $l$ to position $k$, then, without the loss of generality, jobs at positions $k$, $k+1$, \ldots, $l-1$, would be shifted by one to the right. This move causes the change in the set of the worst-case due-dates of at most two elements:
\begin{enumerate}[(i)]
	\item worst-case due-date of job $j$ is equal to $d_j^+$,
	\item worst-case due-date of job $\pi(l)$ (i.e., the job that occupies position $l$ after the shift) is equal to $\min \{ l, d_i^+ \}$; there is no job with due-date equal to $k$ anymore.
\end{enumerate}
Observe that of the due-date of job $j$ does not decrease, which either does not influence the size of independent set $I(\pi)$ (and hence the number $n_a(\pi)$), or increases it at most by one. At the same time, the due-date of job $\pi(l)$ does not increase, which either does not influence the size of independent set $I(\pi)$, or decreases in at most by one. Consequently, scheduling job $j$ at position $k$ on-time does not cause the number $n_a(\pi)$ to increase by more than one, while it does increase the number $n_p(\pi)$ by exactly one. Moving the job $j$ to position $k$ causes the maximum regret of $\pi$ to be no greater than in a schedule with job $j$ on the position $l$.
\qed

\bigskip

Note that the algorithm cannot be used for computing robust solutions for the problem variant with arbitrary weights $w_i$. This is due to the fact that the weight of maximum independent set of $m$ elements can be greater than the weight of maximum independent set of more than $m$ elements, depending on the set of due-dates. Thus the rule of nondecreasing order of $d_j^+$, used for constraining the worst-case due-dates, does not necessarily lead to favorable adversarial solutions in that case.

\section{Experimental Results}\label{sec:experiment}

We have studied the performance of the mixed-integer linear programming formulation \eqref{mip-1:obj}--\eqref{mip-1:13} using randomly generated datasets of varying size. The experiments were designed with the purpose of addressing the following issues:
\begin{enumerate}[1)]
	\item to find out the limiting size of a problem instance (the number of jobs) for which an optimal solution can be found in a reasonable with the use of MIP model, \label{goal:1}
	\item to compare the solution values obtained with the use of proposed model with ones obtained using simpler heuristic algorithms, \label{goal:2}
	\item to determine whether a higher degree of uncertainty in the input data influences the performance of the solution method.
\end{enumerate}

In the first experiment, for each problem size $n$ (the number of jobs) a set of $10$ problem instances was generated. In each instance, one half of all the generated jobs had certain due-dates (i.e., $d_j^-=d_j^+$), randomly uniformly drawn integers between $1$ and $n$. The other half of the set of jobs had uncertain due-dates, with lower bounds being an uniformly drawn integers between $1$ and $n/3$, and interval widths being an uniformly drawn integers between $1$ and $n/3$.

In the second experiment, similarly, for each problem size $n$, a set of $10$ instances was generated. For each job, the lower bound of due-date interval was uniformly drawn integer between $1$ and $n/3$, and the interval width was uniformly drawn integer between $0$ and $\frac{5}{6}n$. Such datasets constitute highly uncertain problem instances, perhaps more than would be expected in practice. However, the goal of this experiment was to test the limits of the solution algorithm. In both experiments for each job the weight was uniformly drawn integer between $1$ and $100$.

In order to evaluate the quality of solutions obtained with the use of MIP we have developed two simple heuristic solution methods, based on removing uncertainty from the input data. Note that since the nominal problem is polynomially solvable, these heuristics run very fast even on large problem instances, thus we omit reporting computation times for them.

The first one, called {\em LB heuristic}, takes interval lower bound of each job, and solves the problem for the obtained scenario using LP \eqref{lp-1:obj}--\eqref{lp-1:2}. This heuristic is equivalent to solving the original problem with the min-max criterion instead of min-max regret. 

The second one, called {\em mid-point (MP) heuristic}, uses interval middle-points of each job as a deterministic scenario, similarly solved by LP \eqref{lp-1:obj}--\eqref{lp-1:2}. This is a standard approach to remove uncertainty used for solving many typical min-max (regret) problems; for a large class of problems it gives 2-approximation algorithm \cite{aissi2009min}. It can be seen as a direct application of stochastic optimization approach to interval uncertainty: a scenario is taken that corresponds to the expected value of the input data with entropy-maximizing probability distribution.

The experiments have been carried out with the use of CPLEX 12.7 software\footnote{The MIP model implementation and the code used in experiments can be obtained from the author on request.}, that implements branch-and-cut type of algorithms, capable of solving to optimality mixed-integer linear programs, as well as producing feasible solutions under specified conditions regarding optimality gap and time limit. The machine used for computations was a Linux system with Intel Xeon processor with 16 cores at 2.3GHz clock rate, and 64 GB of RAM. 

All problem instances in the first experiment were solved to optimality (see Table \ref{tab:1}). It turned out that since only $50\%$ of the generated jobs were uncertain, these instances were easier to solve.

In the second experiment, we have assumed the time limit of $3$ hours for each problem instance. If no solution within $1\%$ of an optimal value has been found within that limit, the computations were aborted, and the best feasible solution has been reported instead. For the latter, as a measure of solution quality we have used the relative integrality gap:
$$
	Q = \frac{UB - LB}{UB} \cdot 100,
$$
where $UB$ is the best upper bound on the optimal robust solution, i.e., the value of the best feasible (integer) solution found, and $UB$ is the best lower bound on the optimal robust solution, i.e., the highest value of the linear programming relaxation used by the branch-and-cut algorithm. 

Since for the highly uncertain problem instances with $50$ and more jobs the MIP was unable to verify optimality within assumed time limit, we have developed a simple decomposition heuristic that allows to handle such problem instances more efficiently. The heuristic decomposes the problem of size $n$ into $m$ subproblems of size about $\lfloor n/m \rfloor$, where $2 \leq m \leq n$. Each subproblem is then solved to optimality with the use of MIP. This gives $m$ schedules that need to be merged into a final schedule of length $n$. Such a schedule can be then used as an initial feasible solution for MIP with full problem data (as a so-called {\em warm-start} solution). This last step can be applied in order to improve the initial solution created by merging smaller schedules, and should be run for a prespecified amount of time (we have set 300 seconds in the experiments).

The summary of the {\em MIP decomposition heuristic} is the following:

\medskip

{\bf Algorithm D.}
\begin{enumerate}
	\item For $i=1, \ldots, m-1$, sample $\lfloor n/m \rfloor$ jobs from $J$, denote them by $J_i$.
	
	\item Denote by $J_m$ the set of remaining jobs.
	
	\item For $i=1, \ldots, m$, solve MIP \eqref{mip-1:obj}--\eqref{mip-1:13} for the set of jobs $J_i$, obtaining schedules $\pi^{(i)}$.
	
	\item Let $\tilde{\pi}$ denote the final schedule for $J$; initialize it with empty $n$-vector.
	\item For $k=1, \ldots, n$:
	
		\begin{enumerate}
			\item Let $V = \{ \pi^{(i)}(0) : \; i = 1,\ldots,m \}$.
			\item Let $j$ be the job with the highest weight among the jobs in $V$.
			\item Remove job $j$ from its schedule $\pi^{(i)}$, and shift all the remaining jobs by 1 to the left.
			\item Schedule the job $j$ in $\tilde{\pi}$ at position $k$.
		\end{enumerate}	
		
	\item Using $\tilde{\pi}$ as a warm-start solution, run MIP solver on $J$ for a prespecified time.
	
\end{enumerate}

Table \ref{tab:2} contains the results for hard datasets (with a high degree of uncertainty). For instances with up to $40$ jobs the mean computation times are given (excluding the time-outed instances), along with the upper bounds on solutions (these are in fact all optimal solutions for $n$ up to $20$), the lower bounds from best LP relaxations, and optimality gaps $Q$. For the larger instances, the statistics of solution values obtained by the MIP decomposition heuristic are compared to the ones obtained using MIP with time limit of 15 minutes (the column labelled ``MIP subopt.''), as well as the other heuristics. Note that the values of $Q$ cannot be obtained from heuristics.

\begin{table}[ht!]
	\caption{\footnotesize Experimental results for the set of problem instances with $50\%$ of uncertain due-dates. In MIP column the reported values correspond to optimal solutions. For each problem size $n$ the mean and standard deviation values are reported from 10 randomly generated problem instances.}\label{tab:1}

	\centering
	{\footnotesize
\begin{tabular}{|ccccc|cc|cc|}
	\hline
        & \multicolumn{4}{c}{MIP} & \multicolumn{2}{|c}{LB heuristic} & \multicolumn{2}{|c|}{MP heuristic} \\
	\hline
	    & \multicolumn{2}{c}{time} & \multicolumn{2}{c}{optimal value} & \multicolumn{2}{|c|}{value} & \multicolumn{2}{c|}{value} \\
	$n$ & mean & max               & mean & std 			   & mean & std                & mean & std \\
	\hline
	10 & 0.05 & 0.07 & 50.5 & 33.92 & 63.9 & 44.34 & 82.5 & 48.24 \\
	15 & 0.15 & 0.27 & 78.9 & 23.25 & 91.1 & 32.27 & 162.1 & 54.75 \\
	20 & 0.29 & 0.47 & 101.2 & 33.41 & 116.6 & 34.26 & 194.3 & 64.97 \\
	25 & 0.70 & 1.44 & 122.0 & 54.61 & 160.2 & 70.76 & 263.6 & 78.73 \\
	30 & 1.20 & 1.60 & 159.2 & 53.56 & 205.7 & 64.63 & 450.4 & 105.84 \\
	35 & 1.85 & 4.71 & 199.6 & 68.42 & 218.0 & 73.14 & 559.8 & 127.69 \\
	40 & 3.10 & 6.38 & 180.7 & 46.35 & 214.4 & 52.84 & 566.6 & 143.68 \\
	50 & 8.23 & 18.66 & 243.1 & 45.44 & 294.6 & 49.74 & 676.2 & 100.17 \\
	60 & 20.03 & 34.95 & 296.3 & 57.16 & 441.2 & 84.92 & 990.4 & 233.07 \\
	70 & 35.17 & 70.41 & 338.3 & 65.16 & 404.9 & 70.52 & 1025.8 & 135.49 \\
	80 & 60.28 & 148.67 & 383.4 & 62.05 & 463.2 & 96.80 & 1380.4 & 182.46 \\
	\hline
\end{tabular}
	}
\end{table}

\begin{table}[h!]
	\caption{\footnotesize Experimental results for the set of highly uncertain problem instances. For MIP the column ``optimal'' contains the number of instances for which optimal solutions have been found within 3 hour time limit; also reported are mean and standard deviation of best solution values (UB), along with best lower bounds (LB) and optimality gap ($Q$). For larger instances results are obtained by the MIP decomposition method. 
	}\label{tab:2}
	
	\centering
	{\scriptsize
\begin{tabular}{|ccccccc|cc|cc|}
	\hline
	 & \multicolumn{6}{c}{MIP} & \multicolumn{2}{|c}{LB heuristic} & \multicolumn{2}{|c|}{MP heuristic} \\
	\hline
	    & time & optimal & \multicolumn{2}{c}{UB} & LB & $Q$ & \multicolumn{2}{c|}{value} & \multicolumn{2}{c|}{value} \\ 
	$n$ & mean & count   & mean & std & mean             & mean & mean & std & mean & std \\
	\hline
	10 & 1.19 & 10/10 & 242.1 & 81.76 & 236.72 & & 278.5 & 80.76 & 315.1 & 99.83 \\
	15 & 6.96 & 10/10 & 287.5 & 51.94 & 277.33 & & 341.3 & 77.19 & 475.9 & 101.70 \\
	20 & 44.88 & 10/10 & 451.4 & 88.79 & 436.72 & & 521.6 & 97.19 & 711.0 & 92.27 \\
	25 & 403.76 & 9/10 & 517.1 & 83.68 & 499.51 & 3.42 & 582.5 & 77.05 & 864.8 & 136.33 \\
	30 & 2868.15 & 9/10 & 621.8 & 66.88 & 591.96 & 4.57 & 733.8 & 107.62 & 1020.7 & 131.39 \\
	35 & 5028.47 & 4/10 & 719.3 & 128.64 & 641.07 & 9.80 & 824.0 & 164.76 & 1244.9 & 169.94 \\
	40 & 2896.61 & 1/10 & 788.0 & 72.01 & 655.76 & 16.3 & 864.4 & 124.72 & 1283.2 & 147.69 \\
	\hline
	    & \multicolumn{3}{c}{decomposition heuristic} & \multicolumn{3}{c|}{MIP subopt.} & \multicolumn{2}{c|}{value} & \multicolumn{2}{c|}{value} \\
	$n$ & mean & std &  & mean & std & $Q$ & mean & std & mean & std \\
	\hline
	50 & 1004.1 & 145.54 & & 1029.3 & 141.83 & 45.77 & 1121.7 & 162.33 & 1783.0 & 218.30 \\
	60 & 1166.7 & 136.38 & & 1223.8 & 113.33 & 60.03 & 1430.9 & 211.46 & 2182.3 & 223.57 \\
	70 & 1406.8 & 136.69 & & 1419.3 & 131.07 & 65.24 & 1496.4 & 136.34 & 2515.3 & 306.88 \\
	80 & 1740.4 & 160.57 & & 1693.2 & 140.71 & 72.23 & 1853.3 & 280.82 & 2840.7 & 418.29 \\
	\hline
\end{tabular}
	}
\end{table}

As it was anticipated, the higher degree of uncertainty caused more difficulties for the proposed solution method. For $30$ and more jobs of this type, in many cases an exact solution was not found within assumed time limit. However, even for them a quite accurate approximations were obtained in reasonable time. The results for larger instances ($n=50$ and more) indicate that MIP solver is able to find near-optimal solutions quickly, but the vast majority of computation time is spent on proving their (near-) optimality (i.e., computing increasingly tighter lower bounds from LP relaxations). Due to this, especially efficient is the proposed decomposition heuristic, which can serve as a warm-start for the MIP solver running for a prespecified amount of time.

In all cases the solutions obtained by solving MIP were superior to both lower bound (min-max) heuristic and mid-point heuristic, while among these two, the former was consistently better. We can conclude that taking the most pessimistic (all lower bounds) scenario usually gives a fairly good solution also in terms of the maximum regret. On the other hand, the mid-point scenario heuristic (and associated stochastic approach), may give very bad solutions with respect to the worst-case scenarios.

\section{Conclusions}\label{sec:conclusions}

We have developed solution algorithms for two variants of single machine scheduling problem to minimize the weighted number of late jobs with interval due-date uncertainty. 
For the case with all unit weights we have described an algorithm running in time $O(n^2)$. For the general case we have formulated a mixed-integer linear program, and validated its efficiency in an experimental study, using problem instances characterized by different amounts of uncertainty. We have also assessed the solution quality of two simple heuristic algorithms, and developed a decomposition method for handling large problem instances.

It is unknown whether the weighted case of this problem is NP-hard or if an exact polynomial time algorithm exists; we conjecture that the problem is NP-hard. The considered problems concern ones of the basic scheduling models with uncertain parameters. There are many practically important extensions, where robust solutions are desired. Such include, but are not limited to, the problem variants with release times, ordering constraints, and parallel machines. We note however that in such restrictive parametric assumptions as the interval uncertainty, very often the computational complexity of such problems increases significantly, in comparison to the counterparts with certain parameters. It is thus interesting to isolate the cases, where solutions can be obtained quickly in practice, regardless of the strong uncertainty assumptions. 

\begin{table}[ht!]
	\caption{List of defined symbols.}\label{tab:0}
	
	\centering
\begin{tabular}{|c|c|}
	\hline
	symbol & description \\
	\hline
	$J$ & set of jobs \\
	$p_i$ & processing time of $i$th job \\
	$w_i$ & weight of $i$th job \\
	$d_i$ & (exact) due-date of $i$th job \\
 	$[d_i^-, d_i^+]$ & uncertainty interval of due-date of $i$th job \\
	$W$ & sum of all weights, i.e., $\sum_{i \in J} w_i$ \\
	$\mathcal{U}$ & set of all scenarios, i.e., $\mathcal{U} = \times_{i=1}^n [d_i^-, d_i^+]$ \\
	$d_i^S$ & due-date of $i$th job in scenario $S \in \mathcal{U}$ \\
	$\mathcal{P}$ & set of all permutations of $J$ \\
	$\pi$ & a schedule (a permutation of $J$) \\
	$\pi^{-1}(i)$ & the time when job $i$ starts in schedule $\pi$ \\
	$C(\pi,i)$  & completion time of job $i$ in schedule $\pi$ \\
	$F(\pi)$ & total weight of late jobs (scheduling objective) \\
	$R(\pi, S)$ & regret of solution $\pi$ in scenario $S \in \mathcal{U}$ \\
	$Z(\pi)$ & maximum regret of solution $\pi$ \\
	
	\hline
\end{tabular}
\end{table}



\end{document}